\documentclass[onecolumn,draftclsnofoot,journal]{IEEEtran}
\usepackage{ascmac}	
\usepackage{amssymb,amsmath,bm}
\usepackage[dvipdfmx]{graphicx}

\newcommand{\pvec}[1]{\stackrel{\leftrightarrow}{#1}}
\newtheorem{theorem}{Theorem}
\newtheorem{definition}{Definition}
\newtheorem{lemma}{Lemma}

\ifCLASSINFOpdf
\else
\fi
\begin{document}
%
\title{Linear Programming Decoding of Binary Linear Codes for Symbol-Pair Read Channels}
%
%
%

\author{Shunsuke Horii, Toshiyasu Matsushima, and Shigeichi Hirasawa }

\maketitle

\begin{abstract}
In this paper, we develop a new decoding algorithm of a binary linear
codes for symbol-pair read channels.
Symbol-pair read channel has recently been introduced by Cassuto and
Blaum to model channels whose write resolution is higher than its read resolution.
The proposed decoding algorithm is based on a linear programming (LP).
It is proved that the proposed LP decoder has the maximum-likelihood (ML)
certificate property, i.e., the output of the decoder is guaranteed to
be the ML codeword when it is integral.
We also introduce the \textit{fractional pair distance} $d_{fp}$ of a code which
is a lower bound on the pair distance.
It is proved that the proposed LP decoder will correct up to $\lceil
d_{fp}/2\rceil-1$ pair errors.
\end{abstract}


%
\IEEEpeerreviewmaketitle

\section{Introduction}
Recent advances in data storage technologies make it
possible to write information to storage devices with high resolution.
However, it causes a problem of the gap between write resolution and
read resolution.
In the case that read resolution is lower than write resolution, it is
difficult to successfully read the individual symbols written to the
storage devices.
The \textit{symbol pair read channel} proposed by Cassuto and Blaum
\cite{Cassuto_CFS_2011} is the channel model that models reading process
of such storage devices.
On each reading operation, two consecutive symbols are read.
We need new error-correction framework for the new channel model.
In the reading process, symbol-pair errors would occur rather than
individual symbol errors, where a symbol-pair error means at least one
of the symbol pair is read erroneously.
The main task is to construct error correcting codes and decoding
algorithms that can correct many symbol-pair errors as possible.

Some error correcting codes and decoding algorithms have been studied.
Cassuto and Blaum \cite{Cassuto_CFS_2011} built a framework for
correcting pair errors, studying a new metric called pair-distance and
its relation to pair error correction capability.
They also provided code constructions and decoding algorithms and
derived some bounds
on the size of optimal symbol-pair codes.
Cassuto and Listsyn \cite{Cassuto_SCA_2011} proposed algebraic
construction of cyclic symbol-pair codes and studied its property.
Yaakobi et al. \cite{Yaakobi_DOC_2012} proposed efficient decoding
algorithms for the cyclic symbol-pair codes.
Chee et al. \cite{Chee_MDS_2012} established a Singleton-type bound for
symbol-pair codes and constructed codes that meet the Singleton-type
bound.
Hirotomo et al. \cite{Hirotomo_SDO_2014} proposed the decoding algorithm
for symbol-pair codes based on the newly defined parity-check matrix and
syndromes.

In this paper, we develop a new decoding algorithm of a binary linear
codes for symbol-pair read channels.
We establish a decoding algorithm which is based on a linear programming
(LP).
The LP based decoding algorithm (LP decoder) of a binary linear codes
for memoryless channels was originally proposed by Feldman et
al. \cite{Feldman_ULP_2005} and studied further by many researchers.
The LP decoder has many attractive properties as follows.
\begin{itemize}
 \item The output of the decoder is guaranteed to be the
	   maximum-likelihood (ML) codeword (ML certificate property) \cite{Feldman_ULP_2005}.
 \item LP decoding with expander codes achieves the capacity of any binary-input memoryless symmetric
	   LLR-bounded (MSB) channel \cite{Feldman_LDA_2005}.
 \item There exist algorithms which efficiently solve the LP \cite{Taghavi_AMF_2008}\cite{Zhang_EIL_2013}.	   
\end{itemize}
It is natural to expect that the decoder based on the LP also
works well for symbol-pair read channel.
Not only establishing the LP decoder for symbol-pair read channel, we
also investigate its property.
We show that the LP decoder has ML certificate property even when the
channel is symbol-pair read channel.
We also introduce the \textit{fractional pair distance} $d_{fp}$ of a
code which is a lower bound on the pair distance.
It is proved that the proposed LP decoder will correct up to $\lceil
d_{fp}/2\rceil$ errors.
We examine the average fractional pair distance of randomly chosen LDPC
codes from an ensemble of Gallager \cite{Gallager_LDPC_1962}.

The rest of the paper is organized as follows.
In Section 2, we introduce basic notation and definitions for symbol-pair
read channels.
In Section 3, we establish the LP decoder for symbol-pair read channels
and in Section 4, some properties of the proposed LP decoder are given.
We conclude the paper in Section 5.
%
%
%
%
\section{Preliminaries}

In this section, we briefly review the model and definition of the
symbol-pair read channel and error-correcting codes.
Throughout this paper, we deal with binary linear codes and let codeword
symbols $\Sigma=\left\{0,1\right\}$.
Let $n$ denotes the code length and $\mathcal{C}\subset \Sigma^{n}$
denotes a binary linear code.
For a length $n$ vector
$\bm{x}=\left(x_{0},\ldots,x_{n-1}\right)\in\Sigma^{n}$, the symbol-pair
read vector of $\bm{x}$, denoted by $\pi(\bm{x})$, is defined as
\begin{align*}
 \pi(\bm{x})=((x_{0},x_{1}),(x_{1},x_{2}),\cdots,(x_{n-2},x_{n-1}),(x_{n-1},x_{0})).
\end{align*}
Unless stated otherwise, all indices are taken modulo $n$.
The pair-distance between vectors $\bm{x}\in\Sigma^{n}$ and
$\bm{y}\in\Sigma^{n}$ is defined as
\begin{align*}
 D_{p}(\bm{x},\bm{y})&=D_{H}(\pi(\bm{x}),\pi(\bm{y}))\\
 &=\left|\left\{i:(x_{i},x_{i+1})\neq (y_{i},y_{i+1})\right\}\right|,
\end{align*}
where $D_{H}$ denotes the usual Hamming distance.
Similarly, the pair weight of $\bm{x}$ is defined as
\begin{align*}
 W_{p}(\bm{x})&=W_{H}(\pi(\bm{x}))\\
 &=\left|\left\{i:(x_{i},x_{i+1})\neq(0,0)\right\}\right|,
\end{align*}
where $W_{H}$ denotes the usual Hamming weight.
We say that symbol-pair error is occurred at $i$-th position when at
least one of two bits $(x_{i},x_{i+1})$ are changed during reading
operation.

We assume the probabilistic model for symbol-pair read channel.
Let $p<\frac{3}{4}$ be the probability that the symbol-pair error occurs and let
\begin{align*}
 p(\bm{b}|\bm{a})=\left\{
\begin{array}{ll}
 1-p&\mbox{if }\bm{a}=\bm{b} \\
 \frac{p}{3}&\mbox{otherwise}
\end{array}
 \right. \bm{a},\bm{b}\in\left\{0,1\right\}^{2}.
\end{align*}
Let $\pvec{\bm{y}}\in \left(\Sigma\times\Sigma\right)^{n}$ be the
received pair vector.
We assume that pair errors occur independently at each position $i$ and 
\begin{align}
 p(\pvec{\bm{y}}|\bm{x})&=p(\pvec{\bm{y}}|\pi(\bm{x}))\nonumber\\
 &=\prod_{i=0}^{n-1}p(\pvec{y}_{i}|(\pi(\bm{x}))_{i}),\label{channelmodel}
\end{align}
where $\pvec{y}_{i}$ and $(\pi(\bm{x}))_{i}$ are the $i$th symbol-pair
of $\pvec{\bm{y}}$ and $\pi(\bm{x})$, respectively.
We call 
$D_{H}(\pi(\bm{x}),\pvec{\bm{y}})$ the number of pair errors when the $\bm{x}$ is the sent codeword and
$\pvec{\bm{y}}$ is the received pair vector.

\section{LP decoding for symbol-pair read channel}

When $\pvec{\bm{y}}$ is the read pair vector, ML decoding problem is
described as
\begin{align}
 \begin{array}{ll}
  \mbox{minimize}& -\ln p(\pvec{\bm{y}}|\bm{x}) \\
  \mbox{s.t.}& \bm{x}\in \mathcal{C}.
 \end{array}\label{OriginalOptimization}
\end{align}
This problem includes the ML decoding problem for memoryless channels
and therefore it is NP-hard in general \cite{Berlekamp_OTI_1978}.
Here, we derive the relaxed problem.

Let $\tau_{i,(a,b)}(\bm{x})$ be the indicator function which represents
whether $(\pi(\bm{x}))_{i}=(a,b)$ or not, i.e.,
\begin{align*}
 \tau_{i,(a,b)}(\bm{x})=\left\{
 \begin{array}{ll}
  1& \mbox{if }(\pi(\bm{x}))_{i}=(a,b)\\
  0& \mbox{otherwise}\label{tau}
 \end{array}
 \right. .
\end{align*}
We also define the map $T: \Sigma^{n}\to \left\{\Sigma^{2}\right\}^{n}$ as $T(\bm{x})=\left(\left(\tau_{i,(a,b)}(\bm{x})\right)_{(a,b)\in\left\{\Sigma^{2}\right\}}\right)_{i=0,\ldots,n-1}$.
Note that $T: \Sigma^{n}\to \left\{\Sigma^{2}\right\}^{n}$ is an
injection and the relations $x_{i}=\tau_{i,(1,0)}(\bm{x})+\tau_{i,(1,1)}(\bm{x})$,
$x_{i+1}=\tau_{i,(0,1)}(\bm{x})+\tau_{i,(1,1)}(\bm{x})$ hold.

From the assumption for the channel, $-\ln p(\pvec{\bm{y}}|\bm{x})$ can
be written as $\sum_{i=0}^{n-1} -\ln
p(\pvec{y}_{i}|\pi(\bm{x})_{i})$.
Given $\pvec{y}_{i}$, $p(\pvec{y}_{i}|(\pi(\bm{x}))_{i})$ is a function of
$(\pi(\bm{x}))_{i}$.
Let $\lambda_{i,(a,b)}=-\ln
p(\pvec{y}_{i}|(a,b)),(a,b)\in\Sigma^{2}$ then,
\begin{align*}
-\ln p(\pvec{y}_{i}|(\pi(\bm{x}))_{i})=\sum_{(a,b)\in \Sigma^{2}}\lambda_{i,(a,b)}\tau_{i,(a,b)}(\bm{x}).
\end{align*}
We define $\bm{\lambda}=(\bm{\lambda}_{0},\cdots,\bm{\lambda}_{n-1})$¡¤
$\bm{\lambda}_{i}=(\lambda_{i,(a,b)})_{(a,b)\in\Sigma^{2}},\
i=0,\cdots,n-1$, and let $\Lambda$ be the map from $\bm{y}$
to $\bm{\lambda}$ i.e., $\bm{\lambda}=\Lambda(\bm{y})$.

Summarizing above, the optimization problem (\ref{OriginalOptimization})
can be converted to the following optimization problem.
\begin{align}
 \begin{array}{ll}
  \mbox{minimize}&
   \left<\bm{\lambda},\bm{\tau}(\bm{x}) \right>\\
  \mbox{s.t.} & \tau_{i,(a,b)}(\bm{x})\in\left\{0,1\right\},\quad i=0,\ldots,n-1,
   (a,b)\in\Sigma^{2}\\
  & \sum_{(a,b)\in\Sigma^{2}}\tau_{i,(a,b)}(\bm{x})=1,\quad i=0,\ldots,n-1\\
  & x_{i}=\tau_{i,(1,0)}(\bm{x})+\tau_{i,(1,1)}(\bm{x}),\quad
   i=0,\ldots,n-1\\
  & x_{i+1}=\tau_{i,(0,1)}(\bm{x})+\tau_{i,(1,1)}(\bm{x}),\quad i=0,\ldots,n-1\\
  & \bm{x}\in\mathcal{C},
 \end{array}
\end{align}
where $\bm{\tau}(\bm{x})=T(\bm{x})$ and $\left<\bm{\lambda},\bm{\tau}(\bm{x})
\right>=\sum_{i=0}^{n-1}\sum_{(a,b)\in\Sigma^{2}}\lambda_{i,(a,b)}\tau_{i,(a,b)}(\bm{x})
$.
Note that the constraints other than $\bm{x}\in\mathcal{C}$ are
redundant constraints that are automatically satisfied if $\bm{x}\in\mathcal{C}$.

We only rewrote the problem (\ref{OriginalOptimization}) in another form
and therefore it is still hard to solve the optimization problem.
Then, we relax the constraints $\tau_{i,(a,b)}\in\left\{0,1\right\}$ to 
$\tau_{i,(a,b)}\in\left[0,1\right]$ and $\bm{x}\in\mathcal{C}$ to
$\bm{x}\in\mathcal{Q}(H)$, where $H$ is the parity check matrix of 
$\mathcal{C}$ and $\mathcal{Q}(H)$ is the fundamental polytope of
$\mathcal{C}$ \cite{Feldman_ULP_2005}.
$\mathcal{Q}(H)$ has the following properties \cite{Feldman_ULP_2005}¡¥
 \begin{itemize}
 \item $\mathcal{C}\subset \mathcal{Q}(H)$
 \item $\mathcal{Q}(H)\cap \left\{0,1\right\}^{n}=\mathcal{C}$
 \item $\bm{x}\in \mathcal{Q}(H)$ can be expressed by
	   $2n+\sum_{j=1}^{m}2^{d_{j}-1}$ inequalities ($m$ is the number of
	   rows of $H$ and $d_{j}$ is the Hamming weight of the $i$th row of
	   $H$)
 \end{itemize}
The relaxed problem is described as follows.
\begin{align}
 \begin{array}{ll}
  \mbox{minimize}&
  \left<\bm{\lambda},\bm{\tau}\right>\\
  \mbox{s.t.} & \tau_{i,(a,b)}\in[0,1],\quad i=0,\ldots,n-1,
   (a,b)\in\Sigma^{2}\\
  & \sum_{(a,b)\in\Sigma^{2}}\tau_{i,(a,b)}=1,\quad i=0,\ldots,n-1\\
  & x_{i}=\tau_{i,(1,0)}+\tau_{i,(1,1)},\quad i=0,\ldots,n-1\\
  & x_{i+1}=\tau_{i,(0,1)}+\tau_{i,(1,1)},\quad i=0,\ldots,n-1\\
  & \bm{x}\in\mathcal{Q}(H).\label{ProposedOptimization}
 \end{array}
\end{align}
Note that
$\bm{\tau}=(\bm{\tau}_{0},\cdots,\bm{\tau}_{n-1}),\bm{\tau}_{i}=(\tau_{i,(a,b)})_{(a,b)\in\Sigma^{2}}$
are the optimization variables that are independently defined from $\bm{x}$.
If we use LDPC code for $\mathcal{C}$, (\ref{ProposedOptimization}) is
solvable in the polynomial order of $n$. 

The LP decoder for symbol-pair read channel consists of the following steps.
\begin{description}
 \item[1.] Calculate $\bm{\lambda}$ from $\pvec{\bm{y}}$.
 \item[2.] Solve the LP (\ref{ProposedOptimization}).
 \item[3.] If the optimal solution $\bm{x}^{*}$ is integral
			($\bm{x}^{*}\in\left\{0,1\right\}^{n}$), output
			$\bm{x}^{*}$ and declare ``error'' otherwise.
\end{description}
As in the case for the memoryless channel, the following theorem holds.

\begin{theorem}
If the LP decoder outputs a codeword, it is guaranteed to be the ML codeword. 
\end{theorem}
\begin{IEEEproof}
 Let $(\bm{x}^{*},\bm{\tau}^{*})$ be the optimal solution of (\ref{ProposedOptimization}).
Its objective value is smaller than or equal to that of other feasible
solution $(\bm{x},\bm{\tau})$.
For any codeword $\bm{x}'\in \mathcal{C}$, it is easily verified that
$(\bm{x}',T(\bm{x}'))$ is a feasible solution for
(\ref{ProposedOptimization}).
In the following, we will confirm that no other feasible solution of the form
$(\bm{x}',\bm{\tau})$ exists.

First, if $\bm{x}'$ is integral, $\bm{\tau}$ must be integral because
assumption that $\bm{\tau}$ is fractional leads to the contradiction.
For example, if $0<\tau_{i,(1,0)}<1$ then $0<\tau_{i,(1,1)}<1$ must holds
from the constraint $\tau_{i,(1,0)}+\tau_{i,(1,1)}=x_{i}$ and assumption
that $x_{i}$ is integral.
Then $\tau_{i,(0,1)}=1-\tau_{i,(1,1)}$ from the constraint
$\tau_{i,(0,1)}+\tau_{i,(1,1)}=x_{i+1}$ and assumption that $x_{i+1}$ is
integral.
Then
$\sum_{(a,b)\in\Sigma^{2}}\tau_{i,(a,b)}\ge\tau_{i,(0,1)}+\tau_{i,(1,0)}+\tau_{i,(1,1)}=1-\tau_{i,(1,1)}+\tau_{i,(1,0)}+\tau_{i,(1,1)}>1$
and it contradicts the constraint
$\sum_{(a,b)\in\Sigma^{2}}\tau_{i,(a,b)}=1$.
We can lead contradiction in other cases similarly.

Further, there doesn't exist integral $\bm{\tau}$ such that
$(\bm{x}',\bm{\tau})$ is a feasible solution of
(\ref{ProposedOptimization}) unless $\bm{\tau}=T(\bm{x}')$ because
$T:\Sigma^{n}\to\left\{\Sigma^{2}\right\}^{n}$ is an injection.

Therefore $(\bm{x}^{*},\bm{\tau}^{*})$ has cost less than ore equal to
feasible point that corresponds to other codeword $\bm{x}'$.
\end{IEEEproof}

We will show later in Lemma \ref{Lemma_vertex} that if $(\bm{x}^{*},\bm{\tau}^{*})$ is a vertex
 of the feasible region of (\ref{ProposedOptimization}) then $\bm{x}^{*}$ is the vertex
of $\mathcal{Q}(H)$.
When the factor graph of the code is cycle-free, $\mathcal{Q}(H)$ has no
fractional vertex \cite{Feldman_ULP_2005}.
Therefore the proposed optimization problem (\ref{ProposedOptimization})
is an exact formulation of the ML decoding problem of the binary linear code
for symbol-pair read channels in the cycle-free case.

\section{Analysis of LP decoding for symbol-pair read channel}
Let $\mbox{Pr}\left[\mbox{err}|\bm{x}\right]$ be the probability that
the LP decoder makes an error, given that $\bm{x}$ was sent.
It can be described as
\begin{align}
 \mbox{Pr}\left[\mbox{err}|\bm{x}\right] =\mbox{Pr}\left\{\exists(\bm{x}',\bm{\tau}')\ \mbox{s.t}\
 (\bm{x}',\bm{\tau}')\mbox{ is a feasible solution of
 }(\ref{ProposedOptimization}),\ \left<\bm{\lambda},\bm{\tau}'\right>\le
 \left<\bm{\lambda},T(\bm{x})\right>\right\}.\label{def_error}
\end{align}
Here, we analyze the performance of LP decoder for symbol-pair read
channels.

\subsection{The All-Zeros Assumption}
It is common to assume that the transmitted codeword is the all-zeros
vector, because the assumption makes it easy to analyze the performance
of the decoder.
Although it is not straight forward to verify that we can assume the
assumption, we will prove that it is valid for the LP decoder for
symbol-pair read channel as it is valid for memoryless channel
\cite{Feldman_ULP_2005}.

\begin{theorem}
 The error probability of the LP decoder for the symbol-pair read channel (\ref{channelmodel}) is independent of the transmitted codeword.\label{allzeros}
\end{theorem}
\begin{IEEEproof}
 See Appendix \ref{proofallzeros}.
\end{IEEEproof}
According to the theorem, we can assume that the transmitted codeword is
all-zeros vector $\bm{0}$.

\subsection{Fractional Pair Distance}
Here, we introduce the concept of \textit{fractional pair distance}
$d_{fp}$ of a code $\mathcal{C}$ whose parity check matrix is $H$.
\begin{definition}
 The fractional pair distance $d_{fp}$ of a code $\mathcal{C}$ whose parity check
 matrix is $H$ is defined as
  \begin{align}
   d_{fp}&=\min_{\bm{x}\in\mathcal{V}_{\mathcal{Q}}\setminus \left\{\bm{0}\right\}}W_{fp}(\bm{x})\\
   &=\min_{\bm{x}\in\mathcal{V}_{\mathcal{Q}}\setminus \left\{\bm{0}\right\}}\sum_{i=0}^{n-1}w_{fp}((\pi(\bm{x}))_{i}),
  \end{align}
 where $\mathcal{V}_{\mathcal{Q}}$ is the set of vertices of the fundamental
 polytope $\mathcal{Q}(H)$ and
 \begin{align}
  w_{fp}((a,b))=\max\left\{a,b\right\}
 \end{align}
\end{definition}
The fractional distance of the code is defined as
$d_{f}=\min_{\bm{x}\in\mathcal{V}_{\mathcal{Q}}\setminus \left\{\bm{0}\right\}}x_{i}$
in \cite{Feldman_ULP_2005} and the fractional pair distance is an
modified notion of it so that it is suitable for the symbol-pair read channel.
As a trivial relation, $d_{f}\le d_{fp}$ holds from the inequality $x_{i}\le
\max(x_{i},x_{i+1})=w_{fp}((\pi(\bm{x}))_{i}$.
The pair distance of a code $\mathcal{C}$ is defined as
$d_{p}=\min_{\bm{x}\in\mathcal{C}\setminus \bm{0}}W_{p}(\bm{x})$ \cite{Cassuto_CFS_2011}.
It holds that $d_{fp}\le d_{p}$ from the fact that $\mathcal{C}\subset
\mathcal{V}_{\mathcal{Q}}$ and $w_{fp}((a,b))\le 1$.

The performance of the LP decoder is characterized by the fractional
pair distance as stated in the following theorem.

\begin{theorem}\label{main_theorem}
 For a code $\mathcal{C}$ with fractional pair distance $d_{fp}^{*}$, the LP
 decoder can correct up to $\lceil d_{fp}^{*}/2\rceil-1$ pair errors
 occurred through the channel (\ref{channelmodel}).
\end{theorem}

Before proving the theorem, we prove the following lemma.
\begin{lemma}\label{Lemma_vertex}
 Let $(\bm{x},\bm{\tau})$ be a vertex of the feasible region of
 (\ref{ProposedOptimization}).
 Then $\bm{x}$ is a vertex of the fundamental polytope $\mathcal{Q}(H)$,
 i.e., $\bm{x}\in\mathcal{V}_{\mathcal{Q}}$.
\end{lemma}
 \begin{IEEEproof}
  See Appendix \ref{proof_vertex}.
 \end{IEEEproof}
We are now ready to prove theorem \ref{main_theorem}.
 \begin{IEEEproof}
  (\textit{Theorem \ref{main_theorem}})
 We can assume that the transmitted codeword is $\bm{0}$ from theorem
 \ref{allzeros}.
 The decoding failure of LP decoder means that the optimal solution
 $(\bm{x}^{*},\bm{\tau}^{*})$ satisfies $\bm{x}^{*}\neq\bm{0}$.
 From the assumption that the code has the fractional pair distance
 $d^{*}_{fp}$ and Lemma \ref{Lemma_vertex}, it holds that $\sum_{i=0}^{n-1}w_{fp}((\pi(\bm{x}^{*}))_{i})\ge
 d_{fp}^{*}$.
 Let $\mathcal{E}=\left\{i\ |\ \pvec{y}_{i}\neq (0,0)\right\}$ be the
 set of positions that pair errors occurred.

To $(\bm{0},\bm{\tau}^{0})$ be a feasible solution of
 (\ref{ProposedOptimization}), $\bm{\tau}^{0}=T(\bm{0})$ must hold and
 $\tau_{i,(0,0)}^{0}=1$,
 $\tau_{i,(0,1)}^{0}=\tau_{i,(1,0)}^{0}=\tau_{i,(1,1)}^{0}=0$ for
 $i=0,\ldots,n-1$.
 Since $(\bm{x}^{*},\bm{\tau}^{*})$ is the optimal solution of
 (\ref{ProposedOptimization}), it holds that
 \begin{align}
    \left<\bm{\lambda},\bm{\tau}^{*}\right>\le \left<\bm{\lambda},\bm{\tau}^{0}\right>.\label{14}
 \end{align}
 Subtracting the right-hand side from the left-hand side in (\ref{14}),
 we obtain
\begin{align}
\left<\bm{\lambda},\bm{\tau}^{*}\right>-\left<\bm{\lambda},\bm{\tau}^{0}\right>&=
 \sum_{i=0}^{n-1}\sum_{(a,b)\in\Sigma^{2}}\lambda_{i,(a,b)}(\tau_{i,(a,b)}^{*}-\tau_{i,(a,b)}^{0})\nonumber\\
 &=\sum_{i\in\mathcal{E}}\sum_{(a,b)\in\Sigma^{2}}\lambda_{i,(a,b)}(\tau^{*}_{i,(a,b)}-\tau_{i,(a,b)}^{0})+\sum_{i\notin\mathcal{E}}\sum_{(a,b)\in\Sigma^{2}}\lambda_{i,(a,b)}(\tau^{*}_{i,(a,b)}-\tau_{i,(a,b)}^{0}).\label{15}
\end{align}
 For $i\in\mathcal{E}$, it holds that
 \begin{align*}
 \lambda_{i,(a,b)}=\left\{
\begin{array}{ll}
 -\ln \frac{p}{3}&\mbox{if }(a,b)=(0,0) \\
 -\ln (1-p)&\mbox{if }(a,b)=\pvec{y}_{i}\\
 -\ln \frac{p}{3}&\mbox{otherwise}.
\end{array}
 \right. 
 \end{align*}
 Therefore the first term in (\ref{15}) can be expanded as
 \begin{align}
&
 \sum_{i\in\mathcal{E}}\left(-\ln\frac{p}{3}(\tau_{i,(0,0)}^{*}-\tau_{i,(0,0)}^{0})-\ln(1-p)(\tau^{*}_{i,\pvec{y}_{i}}-\tau^{0}_{i,\pvec{y}_{i}})-\sum_{(a,b)\in\Sigma^{2}\setminus((0,0)\cup\pvec{y}_{i})}\ln\frac{p}{3}(\tau^{*}_{i,(a,b)}-\tau^{0}_{i,(a,b)})\right)\nonumber\\
 &=\sum_{i\in\mathcal{E}}\left(-\ln\frac{p}{3}(\tau_{i,(0,0)}^{*}-1)-\ln(1-p)\tau^{*}_{i,\pvec{y}_{i}}-\sum_{(a,b)\in\Sigma^{2}\setminus((0,0)\cup\pvec{y}_{i})}\ln\frac{p}{3}\tau^{*}_{i,(a,b)}\right)\nonumber\\
 &=\sum_{i\in\mathcal{E}}\left(-\sum_{(a,b)\in\Sigma^{2}\setminus
 \pvec{y}_{i}}\ln\frac{p}{3}\tau^{*}_{i,(a,b)}-\ln(1-p)\tau^{*}_{i,\pvec{y}_{i}}\right)+|\mathcal{E}|\ln\frac{p}{3}\nonumber\\
&=\sum_{i\in\mathcal{E}}\left(-\ln\frac{p}{3}(1-\tau^{*}_{i,\pvec{y}_{i}})-\ln(1-p)\tau^{*}_{i,\pvec{y}_{i}}\right)+|\mathcal{E}|\ln\frac{p}{3}\nonumber \\
&=\sum_{i\in\mathcal{E}}\left(\ln\frac{p}{3(1-p)}\tau^{*}_{i,\pvec{y}_{i}}\right).
 \end{align}
 For $i\notin \mathcal{E}$, it hods that
 \begin{align}
 \lambda_{i,(a,b)}&=\left\{
\begin{array}{ll}
 -\ln(1-p)&\mbox{if }(a,b)=(0,0) \\
-\ln\frac{p}{3}&\mbox{otherwise}.
\end{array}
\right. 
 \end{align}
 Therefore the second term in (\ref{15}) can be expanded as
 \begin{align}
 &\sum_{i\notin\mathcal{E}}\left(-\ln(1-p)(\tau^{*}_{i,(0,0)}-\tau^{0}_{i,(0,0)})-\sum_{(a,b)\in\Sigma^{2}\setminus
 (0,0)}\ln\frac{p}{3}(\tau^{*}_{i,(a,b)}-\tau^{0}_{i,(a,b)})\right)\nonumber \\
 &=\sum_{i\notin\mathcal{E}}\left(-\ln(1-p)(\tau^{*}_{i,(0,0)}-1)-\sum_{(a,b)\in\Sigma^{2}\setminus(0,0)}\ln\frac{p}{3}\tau^{*}_{i,(a,b)}\right)\nonumber\\
 &=\sum_{i\notin\mathcal{E}}\left(\ln(1-p)\sum_{(a,b)\in\Sigma^{2}\setminus(0,0)}\tau^{*}_{i,(a,b)}-\ln\frac{p}{3}\sum_{(a,b)\in\Sigma^{2}\setminus(0,0)}\tau^{*}_{i,(a,b)}\right)\nonumber\\
 &=\sum_{i\notin\mathcal{E}}\ln\frac{3(1-p)}{p}\sum_{(a,b)\in\Sigma^{2}\setminus(0,0)}\tau^{*}_{i,(a,b)}.
 \end{align}
 Multiplying (\ref{15}) by $\left(\ln\frac{p}{3(1-p)}\right)^{-1}$,
 \footnote{note that $\left(\ln\frac{p}{3(1-p)}\right)^{-1}<0$ when $p<0.75$}
 \begin{align}
\sum_{i\in\mathcal{E}}\tau^{*}_{i:\pvec{y}_{i}}-\sum_{i\notin
 \mathcal{E}}\sum_{(a,b)\in\Sigma^{2}\setminus
 (0,0)}\tau^{*}_{i,(a,b)} &\le |\mathcal{E}|-\sum_{i\notin
 \mathcal{E}}\sum_{(a,b)\in\Sigma^{2}\setminus
 (0,0)}\tau^{*}_{i,(a,b)}\nonumber\\
 &\le |\mathcal{E}|-\sum_{i=0}^{n-1}\sum_{(a,b)\in\Sigma^{2}\setminus
 (0,0)}\tau^{*}_{i,(a,b)}+\sum_{i\in\mathcal{E}}\sum_{(a,b)\in\Sigma^{2}\setminus
 (0,0)}\tau^{*}_{i,(a,b)}\nonumber\\
 &\le 2|\mathcal{E}|-d_{fp}^{*}\nonumber \\
 &<0\label{19},
 \end{align}
 where the first line follows form $\tau_{i:\pvec{y}_{i}}^{*}\le 1$ and
 the third line follows from $\sum_{(a,b)\in\Sigma^{2}\setminus
 (0,0)}\tau_{i,(a,b)}^{*}\le 1$ and
 \begin{align}
   \sum_{(a,b)\in\Sigma^{2}\setminus (0,0)}\tau_{i,(a,b)}^{*}&\ge
 \max\left\{\tau^{*}_{i,(1,0)}+\tau^{*}_{i,(1,1)},\tau^{*}_{i,(0,1)}+\tau^{*}_{i,(1,1)}\right\}\nonumber\\
 &=\max\left\{x_{i}^{*},x_{i+1}^{*}\right\}\nonumber\\
  &=w_{fp}((\pi(\bm{x}^{*}))_{i})\nonumber\\
  &\ge d_{fp}^{*}.
 \end{align}
 The inequality (\ref{19}) suggests that $(\ref{15})>0$ and it
 contradicts (\ref{14}).
 Therefore $\bm{x}^{*}=\bm{0}$.
 \end{IEEEproof}

 \subsection{Computing the Lower Bound of the Fractional Pair Distance}
 In \cite{Feldman_ULP_2005}, the efficient way to compute the fractional
 distance of an LDPC code is presented.
 We can compute the lower bound of the fractional pair distance of an
 LDPC code in a similar way.
 The reason that only lower bound can be computed is explained later.
 To compute the fractional pair distance, we have to compute
 $\min_{\bm{x}\in\mathcal{V}_{\mathcal{Q}}\setminus\left\{\bm{0}\right\}}W_{fp}(\bm{x})$.
 Note that $W_{fp}(\bm{x})$ is a convex function of $\bm{x}$ since it
 can be expressed as the maximum of linear functions.
 Let $\mathcal{F}$ be the set of facets in $\mathcal{Q}(H)$ which
 $\bm{0}$ doesn't sit on.
 We choose a facet in $\mathcal{F}$ and
 minimize $W_{fp}(\bm{x})$ over that facet.
 The obtained minimum is a lower bound of the values of $W_{fp}(\bm{x})$
 of the vertices in the facet.
 The minimum value obtained over all facets in $\mathcal{F}$ is the
 lower bound of $W_{fp}(\bm{x})$ over all vertices other than $\bm{0}$.

 \begin{figure}[t]
  \begin{center}
    \includegraphics[keepaspectratio=true,
   width=\linewidth]{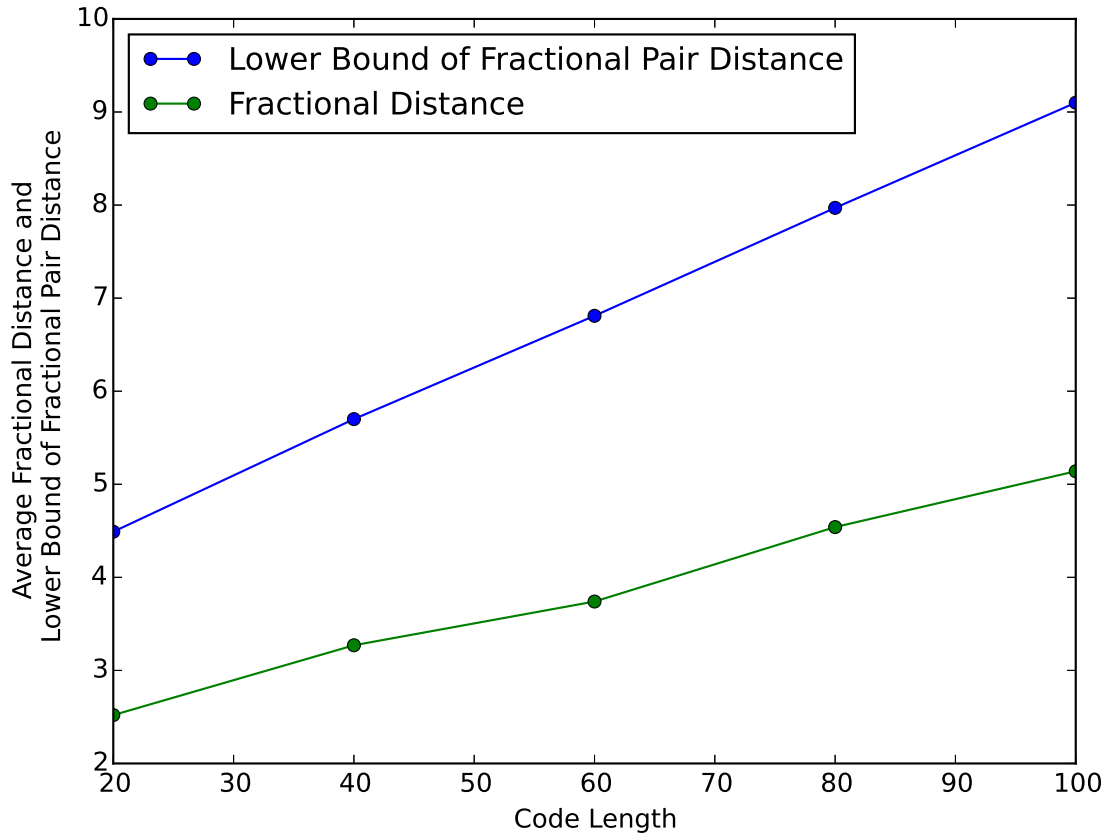}
  \end{center}
    \caption{The average fractional distance $d_{f}$ and the lower bound
  of the fractional pair distance $d_{fp}$ as functions of length for a
  randomly generated LDPC code, with variable degree 3, check degree 4,
  from an ensemble of Gallager \cite{Gallager_LDPC_1962}.}
  \label{fig_fp_distance}
 \end{figure}
 
 Fig. \ref{fig_fp_distance} shows the average fractional pair distance
 of a randomly generated LDPC code.
 The generated LDPC code has variable degree 3, check degree 4, and is
 randomly generated from an ensemble of Gallager \cite{Gallager_LDPC_1962}.
 It seems that the lower bound of the fractional pair distance grows
 linearly with the block length.
 The fractional pair distance is about $1.8$ times as large as the fractional distance.
 Since the fractional pair distance is a lower bound of the pair
 distance, the figure also gives a lower bound of the pair distance of the
 LDPC code.

 \subsection{Numerical Experiments}

\begin{figure}[t]
  \begin{center}
    \includegraphics[keepaspectratio=true,
   width=\linewidth]{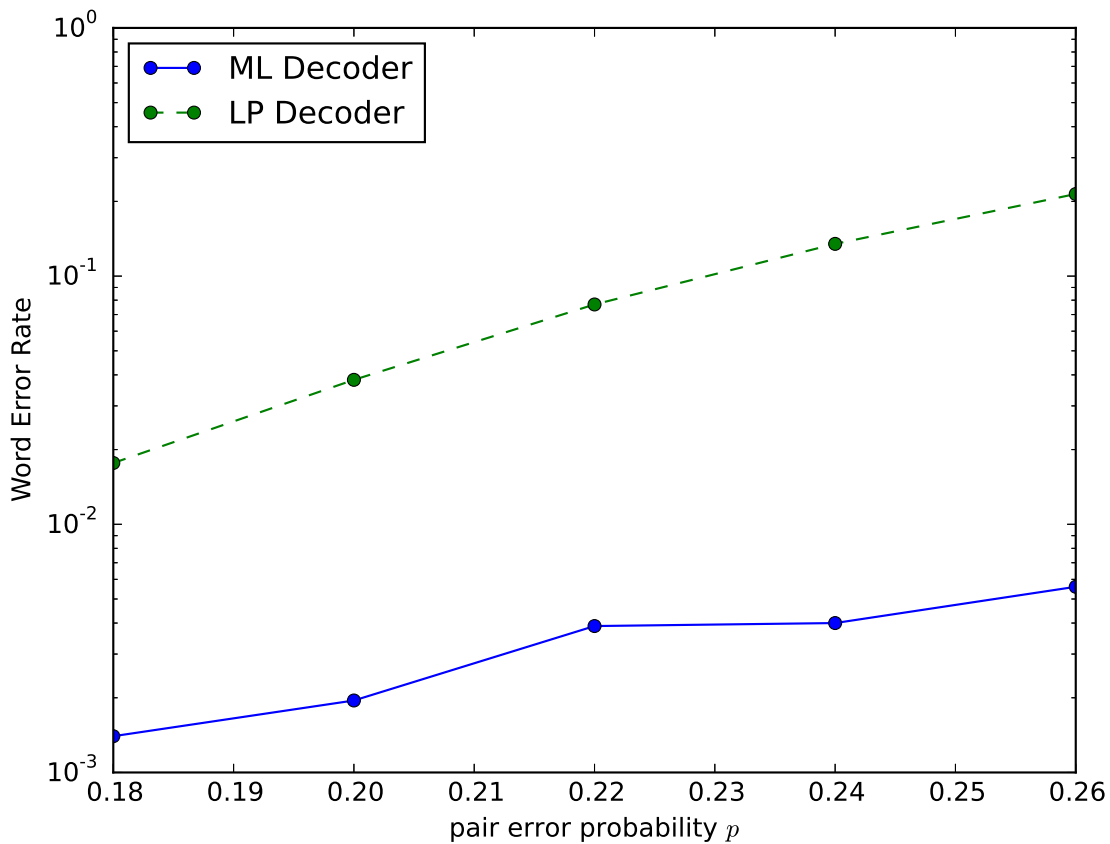}
  \end{center}
\caption{A comparison between the performance of LP decoding and ML
 decoding using the same random rate-1/4 LDPC code with length 60.}
  \label{fig_WordErrorRate}
\end{figure}

 Here, we see the performance of the proposed LP decoder through
numerical experiments.
We used a randomly generated rate-1/4 LDPC code with variable degree 3
and check degree 4.
As one benefit of formulating the decoding problem as an LP problem, we
can obtain ML decoder by solving the corresponding integer linear
program.
Fig. \ref{fig_WordErrorRate} shows an error-rate comparison for a block
length of 60.
There exists a gap between the performance of LP decoder and that of ML
decoder.
We hope that this gap can be filled by tightening the relaxation as in
the case for memoryless channels \cite{Feldman_ULP_2005}.

 \section{Conclusion}
 We have developed an LP based decoding algorithm of binary linear codes
 for symbol-pair read channel.
 We also proved some results on its error-correcting performance.
 We introduced a notion called the fractional pair distance of a code,
 which characterize the lower bound of the number of correctable pair errors that the
 proposed algorithm can correct.

 There are some future directions of the work presented here.
 We assumed that the pair error occurs at each position independently.
 However, this assumption may not hold for some cases.
 The LP decoding for finite-state channels is proposed in
 \cite{Kim_OTJ_2011} and we can consider the symbol-pair read channel
 with finite state.
 To develop the LP decoding algorithm for such channel is a future work.

 In this paper we assumed that the code is binary.
The LP decoding for nonbinary linear codes for memoryless channel was
presented in \cite{Flanagan_LPD_2009}.
It would be interesting to develop the LP decoding algorithm of
nonbinary linear codes for symbol-pair read channel.

\appendices
\section{Proof of theorem \ref{allzeros}} \label{proofallzeros}
In this appendix, we prove theorem \ref{allzeros}.
Recall that $\mbox{Pr}[\mbox{err}|\bm{x}]$ is the probability that the
LP decoder fails, given that $\bm{x}$ was transmitted.
It suffices to show that
\begin{align}
 \mbox{Pr}[\mbox{err}|\bm{x}]=\mbox{Pr}[\mbox{err}|\bm{0}],
\end{align}
for an arbitrary transmitted codeword $\bm{x}\in\mathcal{C}$.
Define $B(\bm{x})$ as follows.
\begin{align}
 B(\bm{x})=\left\{\bm{\pvec{y}}\in(\Sigma^{2})^{n}\ |\ \exists(\bm{x}',\bm{\tau}')\mbox{
 s.t. }(\bm{x}',\bm{\tau}')\mbox{ is a feasible solution of
 (\ref{ProposedOptimization}), } \bm{x}\neq \bm{x}', \left<\bm{\lambda},\bm{\tau}'\right>  \le \left<\bm{\lambda},T(\bm{x})\right>\right\}.
\end{align}
Note that $\bm{\lambda}$ is a function of the received vector
$\bm{\pvec{y}}$.
$B(\bm{x})$ is the set of received vectors $\bm{\pvec{y}}$ that cause
decoding failure when $\bm{x}$ was transmitted.
We can rewrite (\ref{def_error}) as follows.
\begin{align}
 \mbox{Pr}[\mbox{err}|\bm{x}]=\sum_{\bm{\pvec{y}}\in B(\bm{x})}p(\bm{\pvec{y}}|\bm{x}).\label{err_x}
\end{align}
Applying the above to the codeword $\bm{0}$, we obtain
\begin{align}
 \mbox{Pr}[\mbox{err}|\bm{0}]=\sum_{\pvec{\bm{y}}\in
 B(\bm{0})}p(\pvec{\bm{y}}|\bm{0}).\label{err_0}
\end{align}
In a similar manner to the proof in \cite{Feldman_ULP_2005}, we will
show that the space
$(\Sigma^{2})^{n}$ of received vectors can be partitioned into pairs
$(\pvec{\bm{y}},\pvec{\bm{y}}_{\bm{0}})$ such that
\begin{align}
 \mbox{Pr}[\pvec{\bm{y}}|\bm{x}]=\mbox{Pr}[\pvec{\bm{y}}_{\bm{0}}|\bm{0}],\\
 \pvec{\bm{y}}\in B(\bm{x})\Leftrightarrow\ \pvec{\bm{y}}_{\bm{0}}\in B(\bm{0}).
\end{align}
These, along with (\ref{err_x}) and (\ref{err_0}), give
\begin{align*}
 \mbox{Pr}[\mbox{err}|\bm{x}]=\mbox{Pr}[\mbox{err}|\bm{0}]. 
\end{align*}
We partition the space of received vectors as follows.
For an arbitrary received vector $\pvec{\bm{y}}$, define
$\pvec{\bm{y}}_{\bm{0}}$ as follows.
Let $(\pvec{\bm{y}}_{\bm{0}})_{i}$ be
$(\pvec{\bm{y}}_{\bm{0}})_{i}=(\pvec{\bm{y}})_{i}\oplus_{p}(\pi(\bm{x}))_{i}$,
where $\oplus_{p}$ denotes the pairwise exclusive OR, that is,
$(a_{1},b_{1})\oplus_{p}(a_{2},b_{2})=(a_{1}\oplus b_{1},a_{2}\oplus
b_{2})$.
Then $(\pvec{\bm{y}}_{\bm{0}})_{i}=(0,0)$ if and only if
$(\pvec{\bm{y}})_{i}=(\pi(\bm{x}))_{i}$ and therefore,
\begin{align}
  \mbox{Pr}[\pvec{\bm{y}}|\bm{x}]=\mbox{Pr}[\pvec{\bm{y}}_{\bm{0}}|\bm{0}],
\end{align}
holds.
Note that the operation to make $\pvec{\bm{y}}_{0}$ from $\pvec{\bm{y}}$
is its own inverse and it gives a valid partition of $(\Sigma^{2})^{n}$
into pairs.

Now it remains to show that $\pvec{\bm{y}}\in B(\bm{x})$ if and only if
$\pvec{\bm{y}}_{\bm{0}}\in B(\bm{0})$.
Before proving it, we introduce some lemmas.

\begin{lemma}\label{first_lemma}
 Let $(\bm{x}^{f},\bm{\tau}^{f})$ be a feasible solution of
 (\ref{ProposedOptimization}).
 We define the relative solution $(\bm{x}^{r},\bm{\tau}^{r})$ of
 $(\bm{x}^{f},\bm{\tau}^{f})$ to $\bm{x}\in\mathcal{C}$ as follows.
 \begin{align}
    \tau^{r}_{i,(a,b)}&=\tau^{f}_{i,(a,b)\oplus_{p}(\pi(\bm{x}))_{i}},\quad
  i=0,\ldots,n-1, (a,b)\in\Sigma^{2},\\
  x^{r}_{i}&=\tau^{r}_{i,(1,0)}+\tau^{r}_{i,(1,1)},\quad i=0,\ldots,n-1.
 \end{align}
 Then the relative solution $(\bm{x}^{r},\bm{\tau}^{r})$ is a feasible solution of (\ref{ProposedOptimization}).
\end{lemma}
\begin{IEEEproof}
$\tau^{r}_{i,(a,b)}\in[0,1]$ for $i=0,\ldots,n-1, (a,b)\in\Sigma^{2}$ and
 $\sum_{(a,b)\in\Sigma^{2}}\tau^{r}_{i,(a,b)}=1$ for $i=0,\ldots,n-1$ are
 obvious.
 From the definition of the relative solution,
 \begin{align}
  \tau^{r}_{i,(0,1)}+\tau^{r}_{i,(1,1)}&=\tau^{f}_{i,(0,1)\oplus_{p}(\pi(\bm{x}))_{i}}+\tau^{f}_{i,(1,1)\oplus_{p}(\pi(\bm{x}))_{i}}\nonumber\\
  &=\left\{
\begin{array}{ll}
 \tau^{f}_{i,(0,1)}+\tau^{f}_{i,(1,1)}&\mbox{if
  }(\pi(\bm{x}))_{i}=(0,0)\mbox{ or }(1,0) \\
  \tau^{f}_{i,(0,0)}+\tau^{f}_{i,(1,0)}=1-(\tau^{f}_{i,(0,1)}+\tau^{f}_{i,(1,1)})&\mbox{if
  }(\pi(\bm{x}))_{i}=(0,1)\mbox{ or }(1,1) ,
\end{array}
  \right.\\
  x_{i+1}^{r}&=\tau^{r}_{i+1,(1,0)}+\tau^{r}_{i+1,(1,1)}=\tau^{f}_{i+1,(1,0)\oplus_{p}(\pi(\bm{x}))_{i+1}}+\tau^{f}_{i+1,(1,1)\oplus_{p}(\pi(\bm{x}))_{i+1}}\nonumber\\
  &=\left\{
\begin{array}{ll}
\tau^{f}_{i+1,(1,0)}+\tau^{f}_{i,(1,1)}&\mbox{if
  }(\pi(\bm{x}))_{i+1}=(0,0)\mbox{ or }(0,1) \\
  \tau^{f}_{i+1,(0,0)}+\tau^{f}_{i+1,(0,1)}=1-(\tau^{f}_{i+1,(1,0)}+\tau^{f}_{i+1,(1,1)})&\mbox{if
  }(\pi(\bm{x}))_{i+1}=(1,0)\mbox{ or }(1,1) .
\end{array}
  \right. 
 \end{align}
 Considering the fact that the second element of $(\pi(\bm{x}))_{i}$
 equals to the first element of $(\pi(\bm{x}))_{i+1}$, it always holds
 that
 $\tau^{r}_{i,(0,1)}+\tau^{r}_{i,(1,1)}=\tau^{r}_{i+1,(1,0)}+\tau^{r}_{i+1,(1,1)}=x_{i+1}^{r}$
 from $x_{i+1}=\tau^{f}_{i,(0,1)}+\tau^{f}_{i,(1,1)}=\tau^{f}_{i+1,(1,0)}+\tau^{f}_{i+1,(1,1)}$.
 Finally, we show that $\bm{x}^{r}\in\mathcal{Q}(H)$.
 From the definition,
 \begin{align}
  x_{i}^{r}&=\tau_{i,(1,0)}^{r}+\tau_{i,(1,1)}^{r}\nonumber\\
  &=\left\{
\begin{array}{ll}
 \tau_{i,(1,0)}^{f}+\tau_{i,(1,1)}^{f}=x_{i}^{f}&\mbox{if
  }(\pi(\bm{x}))_{i}=(0,0)\mbox{ or }(0,1) \\
 \tau_{i,(0,0)}^{f}+\tau_{i,(0,1)}^{f}=1-x_{i}^{f}&\mbox{if
  }(\pi(\bm{x}))_{i}=(1,0)\mbox{ or }(1,1) 
\end{array}
  \right.\nonumber\\
  &=|x_{i}^{f}-x_{i}|.\label{calc_rel}
 \end{align}
It is proved in \cite{Feldman_ULP_2005} that $\bm{x}^{r}$, whose
elements are calculated from (\ref{calc_rel}), satisfies $\bm{x}^{r}\in\mathcal{Q}(H)$.
\end{IEEEproof}

For the relative solutions, following lemma holds.
\begin{lemma}\label{second_lemma}
 For an arbitrary $\bm{x}\in\mathcal{C}$, an arbitrary feasible solution
 $(\bm{x}^{f},\bm{\tau}^{f})$, and an arbitrary received vector
 $\pvec{\bm{y}}$, it holds that
 \begin{align}
  \left<\bm{\lambda},\bm{\tau}^{f}-\bm{\tau}\right>=\left<\bm{\lambda}^{0},\bm{\tau}^{r}-\bm{\tau}^{0}\right>,
 \end{align}
 where $\bm{\lambda}=\Lambda(\pvec{\bm{y}})$, $\bm{\tau}=T(\bm{x})$,
 $\bm{\lambda}^{0}=\Lambda(\pvec{\bm{y}}_{\bm{0}})$,
 $\bm{\tau}^{0}=T(\bm{0})$, and $\bm{\tau}^{r}$ is the relative solution
 of $\bm{\tau}^{f}$ to $\bm{x}$.
\end{lemma}
\begin{IEEEproof}
 From the definition of $\bm{\lambda}^{0}$, it holds that
 \begin{align}
  \lambda_{i,(a,b)}^{0}&=-\ln p(\pvec{y}^{0}_{i}|(a,b))\nonumber\\
  &=-\ln p(\pvec{y}_{i}\oplus_{p}(\pi(\bm{x}))_{i}|(a,b))\nonumber\\
  &=-\ln p(\pvec{y}_{i}|(a,b)\oplus_{p}(\pi(\bm{x}))_{i})\nonumber\\
  &=\lambda_{i,(a,b)\oplus_{p}(\pi(\bm{x}))_{i}}.
 \end{align}
 Therefore, for $i=0,\ldots,n-1$, it holds that
  \begin{align}
 \sum_{(a,b)\in\Sigma^{2}}\lambda_{i,(a,b)}^{0}\tau^{r}_{i,(a,b)}&=\sum_{(a,b)\in\Sigma^{2}}\lambda_{i,(a,b)\oplus_{p}(\pi(\bm{x}))_{i}}\tau^{f}_{i,(a,b)\oplus_{p}(\pi(\bm{x}))_{i}}\\
 &=\sum_{(a,b)\in\Sigma^{2}}\lambda_{i,(a,b)}\tau^{f}_{i,(a,b)},
  \end{align}
 and
 $\left<\bm{\lambda},\bm{\tau}^{f}\right>=\left<\bm{\lambda}^{0},\bm{\tau}^{r}\right>$.
 Further, from
 $\sum_{(a,b)\in\Sigma^{2}}\lambda_{i,(a,b)}\tau_{i,(a,b)}=\lambda_{i,(\pi(\bm{x}))_{i}}$,
 $\sum_{(a,b)\in\Sigma^{2}}\lambda^{0}_{i,(a,b)}\tau^{0}_{i,(a,b)}=\lambda^{0}_{i,(0,0)}$
 and $\lambda^{0}_{i,(0,0)}=\lambda_{i,(\pi(\bm{x}))_{i}}$, it also
 holds that $\left<\bm{\lambda},\bm{\tau}\right>=\left<\bm{\lambda}^{0},\bm{\tau}^{0}\right>$.
\end{IEEEproof}

We can prove the following lemma from Lemma \ref{first_lemma} and Lemma
\ref{second_lemma}.
\begin{lemma}\label{third_lemma}
 Fix a codeword $\bm{x}\in\mathcal{C}$.
 For an arbitrary feasible solution $(\bm{x}^{f},\bm{\tau}^{f})$ of
 (\ref{ProposedOptimization}), $\bm{x}^{f}\neq\bm{x}$, there exists
 another feasible solution $(\bm{x}^{r},\bm{\tau}^{r})$ such that
 $\bm{x}^{r}\neq\bm{0}$ and
 \begin{align}
     \left<\bm{\lambda},\bm{\tau}^{f}-\bm{\tau}\right> =\left<\bm{\lambda}^{0},\bm{\tau}^{r}-\bm{\tau}^{0} \right>.
 \end{align}
 Further, for an arbitrary feasible solution
 $(\bm{x}^{r},\bm{\tau}^{r})$ of (\ref{ProposedOptimization}), $\bm{x}^{r}\neq\bm{0}$, there exists
 another feasible solution $(\bm{x}^{f},\bm{\tau}^{f})$ such that
 $\bm{x}^{f}\neq \bm{x}$ and
 \begin{align}
  \left<\bm{\lambda}^{0},\bm{\tau}^{r}-\bm{\tau}^{0} \right>=\left<\bm{\lambda},\bm{\tau}^{f}-\bm{\tau}\right> .
 \end{align}
\end{lemma}
\begin{IEEEproof}
Let $(\bm{x}^{r},\bm{\tau}^{r})$ be the relative solution of
 $(\bm{x}^{f},\bm{\tau}^{f})$ to $\bm{x}$.
 From Lemma \ref{first_lemma}, $(\bm{x}^{r},\bm{\tau}^{r})$ is a
 feasible solution of (\ref{ProposedOptimization}).
 From the definition, $\bm{x}^{r}\neq\bm{0}$ and applying Lemma
 \ref{second_lemma}, we obtain
 \begin{align}
  \left<\bm{\lambda},\bm{\tau}^{f}-\bm{\tau}\right> =\left<\bm{\lambda}^{0},\bm{\tau}^{r}-\bm{\tau}^{0} \right>.
 \end{align}
 The latter of the lemma can be proved by replacing the role of
 $(\bm{x}^{f},\bm{\tau}^{f})$ and $(\bm{x}^{r},\bm{\tau}^{r})$.
\end{IEEEproof}

We are now ready to prove that $\pvec{\bm{y}}\in B(\bm{x})$ if and only
if $\pvec{\bm{y}}_{\bm{0}}\in B(\bm{0})$ and it completes the proof of
the Theorem \ref{allzeros}.
Assume that $\pvec{\bm{y}}\in B(\bm{x})$.
From the definition of $B(\bm{x})$, there exists a feasible solution $(\bm{x}^{f},\bm{\tau}^{f})$
of
(\ref{ProposedOptimization}) that satisfies $\bm{x}^{f}\neq\bm{x}$ and
\begin{align}
 \left<\bm{\lambda},\bm{\tau}^{f}-\bm{\tau}\right>\le 0.
\end{align}
From Lemma \ref{third_lemma}, there exists a feasible solution
$(\bm{x}^{r},\bm{\tau}^{r})$ such that $\bm{x}^{r}\neq\bm{0}$ and
\begin{align}
 \left<\bm{\lambda}^{0},\bm{\tau}^{r}-\bm{\tau}^{0}\right>\le 0.
\end{align}
It means $\pvec{\bm{y}}_{\bm{0}}\in B(\bm{0})$.
The other direction can be proved with the symmetric argument.

\section{Proof of Lemma \ref{Lemma_vertex}}\label{proof_vertex}
 It is known that a vertex of a polytope of dimension $N$ is uniquely
 determined by $N$ linearly independent facets of the polytope which the
 vertex sits on \cite{Schirijver_TOL_1987}.
 The dimension of the feasible region of (\ref{ProposedOptimization}) is
 $5n$.
 Let $(\bm{x},\bm{\tau})$ be an arbitrary vertex of the feasible region
 of (\ref{ProposedOptimization}) and hence it sits on $5n$ linearly
 independent facets in (\ref{ProposedOptimization}).
We will show that $(\bm{x},\bm{\tau})$ sits on at most $4n$ linearly
 independent facets in (\ref{ProposedOptimization}) which are not
 included in $\bm{x}\in\mathcal{Q}(H)$.
 It leads that $\bm{x}$ sits on at least $n$ linearly independent facets in
 $\bm{x}\in\mathcal{Q}(H)$.
 Since $\bm{x}$ satisfies at most $n$ linearly independent equality,
 $\bm{x}$ sits on just $n$ linearly independent facets in
 $\bm{x}\in\mathcal{Q}(H)$ and therefore
 $\bm{x}\in\mathcal{V}_{\mathcal{Q}}$.

 Here we show that an arbitrary vertex $(\bm{x},\bm{\tau})$ of the
 feasible region of (\ref{ProposedOptimization}) sits on at most $4n$
 linearly independent facets in (\ref{ProposedOptimization}) other than
 the constraints in $\bm{x}\in\mathcal{Q}(H)$.
 Let $E_{1},E_{2},E_{3},E_{4}$ be the sets which are defined as
 \begin{align}
  E_{1}&=\left\{i\in\left\{1,\ldots,n\right\}\ |\ x_{i}\mbox{ and
  }x_{i+1} \mbox{ are integral}\right\},\\
  E_{2}&=\left\{i\in\left\{1,\ldots,n\right\}\ |\ x_{i}\mbox{ is
  integral, }x_{i+1} \mbox{ is not integral}\right\},\\
  E_{3}&=\left\{i\in\left\{1,\ldots,n\right\}\ |\ x_{i}\mbox{ is
  not integral, }x_{i+1} \mbox{ is integral}\right\},\\
  E_{4}&=\left\{i\in\left\{1,\ldots,n\right\}\ |\ x_{i}\mbox{ and
  }x_{i+1} \mbox{ are not integral}\right\}.
 \end{align}
 We show that for any $i\in E_{j},j=1,\ldots,4$,
 $\left\{\tau_{i,(a,b)}\right\}_{(a,b)\in\Sigma^{2}}$, $x_{i}$, and
 $x_{i+1}$ satisfy at most 4 linearly independent constraints in
 (\ref{ProposedOptimization}) other than the constraints in $\bm{x}\in\mathcal{Q}(H)$.
 
 In the case $i\in E_{1}$, $\tau_{i,(x_{i},x_{i+1})}=1$ and
 $\tau_{i,(a,b)}=0, (a,b)\neq(x_{i},x_{i+1})$. Other constraints are not
 linearly independent from these constraints and
 $x_{i},x_{i+1}\in\left\{0,1\right\}$.
 Therefore just 4 linearly independent equality constraints
 $\tau_{i,(a,b)}\in\left\{0,1\right\}, (a,b)\in\Sigma^{2}$ are
 satisfied.

 Next, we consider the case $i\in E_{2}$.
 First we assume that $x_{i}=0$.
 Then $\tau_{i,(1,0)}=\tau_{i,(1,1)}=0$ holds from
 the constraint $x_{i}=\tau_{i,(1,0)}+\tau_{i,(1,1)}$.
 In this case, it holds that $0<\tau_{i,(0,0)},\tau_{i,(0,1)}<1$
 otherwise it leads contradiction to the assumption $x_{i+1}$ is not
 integral (For example, if $\tau_{i,(0,0)}=0$ then $\tau_{i,(0,1)}=1$
 and $x_{i+1}=\tau_{i,(0,1)}+\tau_{i,(1,1)}=1$).
 The equality $x_{i}=\tau_{i,(1,0)}+\tau_{i,(1,1)}$ is linearly
 dependent to the equalities $\tau_{i,(1,0)}=0$, $\tau_{i,(1,1)}=0$ and
 $x_{i}=0$.
 Therefore just 4 linearly independent equality constraints
 $\tau_{i,(1,0)}=0$, $\tau_{i,(1,1)}=0$,
 $\sum_{(a,b)\in\Sigma^{2}}\tau_{i,(a,b)}=1$,
 $x_{i+1}=\tau_{i,(0,1)}+\tau_{i,(1,1)}$ are satisfied.
 When $x_{i}=1$ then $\tau_{i,(0,0)}=\tau_{i,(0,1)}=0$ holds from the
constraints $x_{i}=\tau_{i,(1,0)}+\tau_{i,(1,1)}$ and
 $\sum_{(a,b)\in\Sigma^{2}}\tau_{i,(a,b)}=1$.
 It can be shown that just 4 linearly independent equality constraints
 are satisfied in a similar way for the case $x_{i}=0$.

 For the case $i\in E_{3}$, we can prove that just 4 linearly
 independent equality constraints are satisfied in a similar way for the
 case $i\in E_{2}$ by exchanging the role of $x_{i}$ and $x_{i+1}$.

 Finally, we consider the case $i\in E_{4}$.
 From the assumption that $0<x_{i}<1$ and constraint
 $x_{i}=\tau_{i,(1,0)}+\tau_{i,(1,1)}$, it turned out that one of the three cases
a) $\tau_{i,(1,0)}=0,0<\tau_{i,(1,1)}<1$ or b) $\tau_{i,(1,1)}=0,0<\tau_{i,(1,0)}<1$ or
c) $0<\tau_{i,(1,0)},\tau_{i,(1,1)}<1$ occurs.
\begin{description}
 \item[a)] We consider the case that $\tau_{i,(1,0)}=0,
			0<\tau_{i,(1,1)}<1$ hold.
			It holds that
 $\tau_{i,(0,0)}+\tau_{i,(0,1)}=1-x_{i}$ from the constraints
			$x_{i}=\tau_{i,(1,0)}+\tau_{i,(1,1)}$ and $\sum_{(a,b)\in
			\Sigma^{2}}\tau_{i,(a,b)}=1$.
			One of the three
 cases a-1) $\tau_{i,(0,0)}=0$ or a-2) $\tau_{i,(0,1)}=0$ or a-3)
			$0<\tau_{i,(0,0)},\tau_{i,(0,1)}<1$ occurs.
			\begin{description}
			 \item[a-1)] If we assume $\tau_{i,(0,0)}=0$ then $\tau_{i,(0,1)}=1-x_{i}$ and
						$x_{i+1}=\tau_{i,(0,1)}+\tau_{i,(1,1)}=1$.
						But it contradicts the assumption $x_{i}$ is not integral and it never
						happens.
			 \item[a-2)]  We assume $\tau_{i,(0,1)}=0$. From the
						constraint $x_{i}=\tau_{i,(1,0)}+\tau_{i,(1,1)}$
						and the assumption $\tau_{i,(1,0)}=0$,
						$\tau_{i,(1,1)}=x_{i}$.
						Therefore,
						$x_{i+1}=\tau_{i,(0,1)}+\tau_{i,(1,1)}=0+x_{i}=x_{i}$
						holds.
						Above argument shows that the constraint
						$x_{i+1}=\tau_{i,(0,1)}+\tau_{i,(1,1)}$ is
						linearly dependent to the constraints
						$\tau_{i,(1,0)}=0$, $\tau_{i,(0,1)}=0$,
						$x_{i}=\tau_{i,(1,0)}+\tau_{i,(1,1)}$ and just 4
						linearly independent constraints
						$\tau_{i,(1,0)}=0$, $\tau_{i,(0,1)}=0$,
						$\sum_{(a,b)\in\Sigma^{2}}\tau_{i,(a,b)}=1$,
						$x_{i}=\tau_{i,(1,0)}+\tau_{i,(1,1)}$ are
						satisfied.
			 \item[a-3)]  When $0<\tau_{i,(0,0)},\tau_{i,(0,1)}<1$ holds, 4 linearly independent
						constraints $\tau_{i,(1,0)}=0$,\\
						$\sum_{(a,b)\in \Sigma^{2}}\tau_{i,(a,b)}=1$, $x_{i}=\tau_{i,(1,0)}+\tau_{i,(1,1)}$,
						$x_{i+1}=\tau_{i,(0,1)}+\tau_{i,(1,1)}$ are satisfied.			
			\end{description}
 \item[b)] We consider the case that $\tau_{i,(1,1)}=0$,
			$0<\tau_{i,(1,0)}<1$ hold.
			As in the case a), it holds that
			$\tau_{i,(0,0)}+\tau_{i,(0,1)}=1-x_{i}$ and one of the three
			cases b-1) $\tau_{i,(0,0)}=0$ or b-2) $\tau_{i,(0,1)}=0$ or b-3)
			$0<\tau_{i,(0,0)},\tau_{i,(0,1)}<1$ occurs.
			\begin{description}
			 \item[b-1)] Consider the case that $\tau_{i,(0,0)}=0$
						holds.
						Then $\tau_{i,(0,1)}=1-x_{i}$ and
						$x_{i+1}=\tau_{i,(0,1)}+\tau_{i,(1,1)}=1-x_{i}$
						holds.
						Above argument shows that the constraint
						$x_{i+1}=\tau_{i,(0,1)}+\tau_{i,(1,1)}$ is
						linearly dependent to the constraints
						$\tau_{i,(0,0)}=0$, $\tau_{i,(1,1)}=0$ and
						$x_{i}=\tau_{i,(1,0)}+\tau_{i,(1,1)}$.
						Therefore just 4 linearly independent
						constraints $\tau_{i,(0,0)}=0$,
						$\tau_{i,(1,1)}=0$,
						$\sum_{(a,b)\in\Sigma^{2}}\tau_{i,(a,b)}=1$ and
						$x_{i}=\tau_{i,(1,0)}+\tau_{i,(1,1)}$ are
						satisfied.
			 \item[b-2)] If we assume that $\tau_{i,(0,1)}=0$ holds then
						$x_{i+1}=\tau_{i,(0,1)}+\tau_{i,(1,1)}=0$, but
						it contradict the assumption $x_{i+1}$ is not
						integral.
						Therefore this case never occurs.
			\item[b-3)] When $0<\tau_{i,(0,0)},\tau_{i,(0,1)}<1$ holds, 4 linearly independent
						constraints $\tau_{i,(1,1)}=0$,\\
						$\sum_{(a,b)\in \Sigma^{2}}\tau_{i,(a,b)}=1$, $x_{i}=\tau_{i,(1,0)}+\tau_{i,(1,1)}$,
						$x_{i+1}=\tau_{i,(0,1)}+\tau_{i,(1,1)}$ are satisfied.			
			\end{description}
 \item[c)] Assume that $0<\tau_{i,(1,0)},\tau_{i,(1,1)}<1$ holds.
		   As in the other cases, it holds that $\tau_{i,(0,0)}+\tau_{i,(0,1)}=1-x_{i}$ and one of the three
			cases c-1) $\tau_{i,(0,0)}=0$ or c-2) $\tau_{i,(0,1)}=0$ or c-3)
			$0<\tau_{i,(0,0)},\tau_{i,(0,1)}<1$ occurs.
			\begin{description}
			 \item[c-1)] When $\tau_{i,(0,0)}=0$ holds then
						$\tau_{i,(0,1)}=1-x_{i}$ and therefore
						$0<\tau_{i,(0,1)}<1$.
						In this case, 4 linearly independent constraints
						$\tau_{i,(0,0)}=0$,
						$\sum_{(a,b)\in\Sigma^{2}}\tau_{i,(a,b)}=1$,
						$x_{i}=\tau_{i,(1,0)}+\tau_{i,(1,1)}$ and
						$x_{i+1}=\tau_{i,(0,1)}+\tau_{i,(1,1)}$ are
						satisfied.
			 \item[c-2)] When $\tau_{i,(0,1)}=0$ holds then
						$\tau_{i,(0,0)}=1-x_{i}$ and
						$0<\tau_{i,(0,0)}<1$ hold.
						Therefore 4 linearly independent constraints $\tau_{i,(0,1)}=0$,
						$\sum_{(a,b)\in\Sigma^{2}}\tau_{i,(a,b)}=1$,
						$x_{i}=\tau_{i,(1,0)}+\tau_{i,(1,1)}$ and
						$x_{i+1}=\tau_{i,(0,1)}+\tau_{i,(1,1)}$ are
						satisfied.
			\item[c-3)] Assume that $0<\tau_{i,(0,0)},\tau_{i,(0,1)}<1$
					   holds.
					   In this case, 3 linearly independent constraints
					   $\sum_{(a,b)\in\Sigma^{2}}\tau_{i,(a,b)}=1$,
					   $x_{i}=\tau_{i,(1,0)}+\tau_{i,(1,1)}$ and
					   $x_{i+1}=\tau_{i,(0,1)}+\tau_{i,(1,1)}$ are
					   satisfied.
			\end{description}
\end{description}

\ifCLASSOPTIONcaptionsoff
  \newpage
\fi

\end{document}